# Interactions between π-conjugated chromophores in a giant molecular spoked wheel

D. Würsch,[a] R. May,[*a] G. Wiederer,[a] S.-S. Jester,[b] S. Höger,[b] J. Vogelsang[a] and J. M. Lupton[a]

**We discuss the intriguing photophysics of a giant molecular spoked wheel of π-conjugated arylene-alkynylene chromophores on the single-molecule level. This "molecular mesoscopic" structure, $C_{1878}H_{2682}$, shows fast switching between the 12 identical chromophores since the fluorescence is unpolarised but only one chromophore emits at a time.**

Given that semiconducting organic molecules have gained utilisation in technical applications such as OLEDs and photovoltaics, it is crucial to persist in striving for a profound understanding of the microscopic details of light-matter interaction in these materials.[1–3] The fate of excited states regarding excitation energy transfer pathways between chromophores in these materials is of particular interest. In exciton-exciton annihilation, one exciton is deexcited by transferring its energy to another exciton. For example, photon antibunching (PA), the temporal separation of photons emitted by a single molecule, is an observable result of exciton-exciton annihilation in multichromophoric aggregates, and can give insight into the interchromophoric energy transfer processes.[4,5] It is as yet unknown up to what degree of spatial extension this process can occur.

The measurement of PA (*i.e.* the emission of only one photon at a given time) has now been well established[6] and PA has been observed for isolated single-chromophore molecules[7–9] and multichromophoric molecules or aggregates.[4,5,10] First signatures of this non-classical fluorescence emission emerged in the collective behaviour of multiple chromophores in emission, *e.g.* in the formation of collective on-/off-states or single-step bleaching,[11–13] and enhanced techniques in single-molecule spectroscopy have now revealed a multitude of examples of PA in multichromophoric molecules.[5,14–17] PA is even observed in large conjugated polymer chains, depending on the molecular morphology. For example, unfolded chains of poly(3-hexylthiophene) (P3HT) do not show any PA, but in a folded conformation the polymer shows perfect PA, *i.e.* only one of the many chromophores on the chain emits at a time.[18,19] Related behaviour regarding the photophysical properties has been observed for poly(phenylene-vinyle) derivatives,[20,21] polyfluorene,[22] small conjugated macrocycles[23] and aggregates consisting of multiple poly(phenylene-ethynylene) chains.[10] PA was also demonstrated for biologically relevant molecules, for example for B-phycoerythrin, a protein found in light-harvesting structures in red algae which contains 34 bilin chromophores,[24] and recently in the light-harvesting complex 2 (LH2) extracted from purple bacteria.[25] LH2 is a helical chromophore-protein complex, coordinating 27 bacteriochlorophylls-*a* in the shape of two telescopic cylinders with diameters of 1.8 nm and 3.4 nm, respectively.[26,27] Altogether, the easily observable effect of PA can serve as a powerful tool to provide insight into the intramolecular energy transfer properties of large π-conjugated molecules or aggregates thereof.

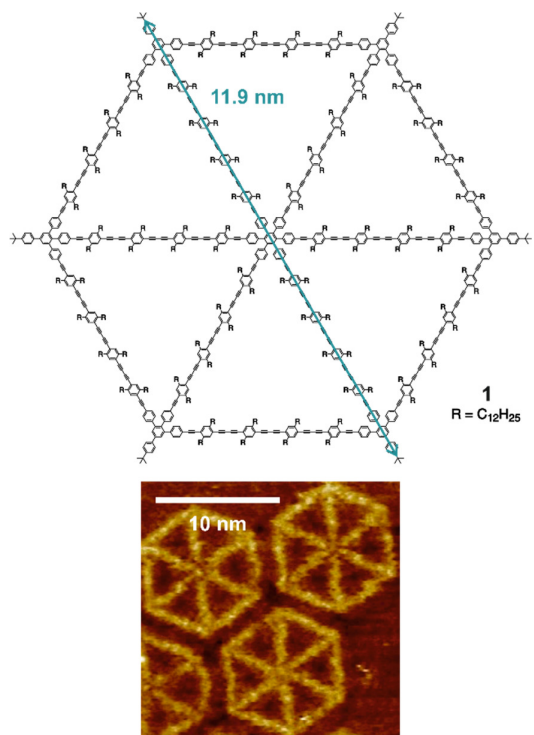

*Figure 1 Chemical structure of the molecular spoked wheel and scanning tunnelling microscopy (STM) image showing a self-assembled monolayer at the solid-liquid interface of highly oriented pyrolytic graphite and octanoic acid ($c = 10^{-7}$ M, $V_s$ = -0.8 V, $I_t$ = 3 pA).*

[a]*Institut für Experimentelle und Angewandte Physik, Universität Regensburg, Universitätsstraße 31, 93053 Regensburg, Germany*

[b]*Kekulé-Institut für Organische Chemie und Biochemie, Universität Bonn, Gerhard-Domagk-Straße 1, 53121 Bonn, Germany*

[*]) *Corresponding author. Email: robert.may@ur.de*

Mimicking the defined spatial arrangement of the chromophores in biological light-harvesting structures and thus avoiding the conformational uncertainty of polymers, spoked-wheel-shaped chromophore arrangements (molecular spoked wheels, MSWs) have been synthesised exhibiting well-defined lateral dimensions and chromophore coordinates.[28,29] Here, we investigate the photophysics of **1**, the largest 12-chromophore-spoked wheel (12 nm in diameter) synthesised to date, having the weight of a polymer or small protein ($C_{1878}H_{2682}$, $M_w$=25 kDa, figure 1). In these MSWs chromophores themselves are integral parts of the structure, thus making additional functionalisation with fluorescence probes unnecessary.

The MSW is designed to have 6 chromophores as spokes and 6 chromophores as the rim, all of identical chemical structure and length and connected to each other by phenylene moieties. The *m*-phenylene connections at the corners of the rim as well as the nonplanar arrangement of the *o/p*-phenylene units at both corners and hub effectively interrupt the conjugation of the linear spoke/rim units, thus leading to 12 independent absorbers.[28] The MSW was synthesised and characterised as described previously.[30] The shape and dimensions of the compound are best visualised by scanning-tunnelling microscopy (STM) imaging of self-assembled monolayers on a highly oriented pyrolytic graphite (HOPG) substrate, and a representative STM image is shown in figure 1.

To investigate the spectroscopic properties of the molecule, we performed single-molecule fluorescence spectroscopy. The sample preparation as well as the measurements followed the methods described in detail previously.[23] In brief, **1** was dispersed in a 1% w/w PMMA/toluene solution and spin-coated onto glass coverslips, typically yielding densities of 40 to 60 sample molecules per 20 x 20 µm$^2$ scan area. Fluorescence transients of single-molecule emission were recorded in an inverted confocal microscope setup. The microscope was equipped with an oil-immersion objective (numerical aperture NA = 1.35), pulsed excitation at 405 nm with a 10 MHz repetition rate was used, and the laser polarisation was switched by 90° every 400 µs. The excitation power was set to approximately 150 Wcm$^{-2}$. As previously observed for similar systems, the presence of oxygen in ambient air was found to be crucial for the measurements to quench non-emissive triplet excited states, thereby raising the rate of photons emitted by each molecule. Triplets will render the individual chromophores non-emissive, but can also quench the fluorescence from other chromophores in the MSW through the process of singlet-triplet annihilation.[31]

The inset in figure 2 shows an ensemble fluorescence spectrum of the MSWs in toluene solution (black curve). The electronic transition peaks at 425 nm and is followed by a pronounced vibronic peak at 450 nm. This emission is close to the UV spectral region, making single-molecule spectroscopy on these compounds challenging. The photoluminescence lifetime of the MSW is found to be ~0.6 ns, which is identical to that of an isolated oligomer unit and implies that the arrangement in the MSW does not induce any substantial modification of the excited state.

An example of a representative photoluminescence (PL) intensity transient of one molecule is depicted in figure 2 together with the corresponding trace of the linear dichroism (*LD*) in emission, for which the PL was split by a polarising beam splitter. Intensities (*I*) for vertically (V) and horizontally (H) polarised fluorescence are recorded, so that in total four detection channels exist, and *LD* is calculated according to $LD=(I_V-I_H)/(I_V+I_H)$. Consecutive bleaching of the multiple chromophores is visible in the stepwise decrease in photon count rate. The presence of multiple chromophores with different orientations in the MSW is also clearly demonstrated through the polarisation anisotropy in emission,[32] which was measured in terms of *LD* of multiple molecules as indicated by the scheme in figure 3. *LD* can be assessed for excitation and emission, respectively. For both *LD* measurements, which are performed simultaneously, the laser polarisation is switched periodically by 90° and the fluorescence is passed through a polarising beam splitter. The inset in figure 3 displays a fluorescence microscope scan image with the single molecules shown in false colours depending on the respective dominant channel of molecular polarisation (the four possible combinations are: H excitation, H detection: blue; H excitation, V detection: green; V excitation, H detection: yellow; V excitation, H detection: red). Mixtures of polarisation channels appear as superposition colours. *LD* values were extracted from such confocal scan images with a software. The *LD* value of each molecule was extracted only during the first approx. 250 ms of excitation so that bleaching effects can be neglected. The *LD* ranges from -1 to +1. Since the orientation of an absorbing and emitting dipole is arbitrary with respect to the laser polarisation and the polarising beam splitter, one has to consider the complete distribution of *LD* values obtained from multiple single molecules. Linearly polarised emitters and absorbers tend to yield large *LD* values of ±1, whereas unpolarised transitions will result in a grouping of *LD* values around 0.

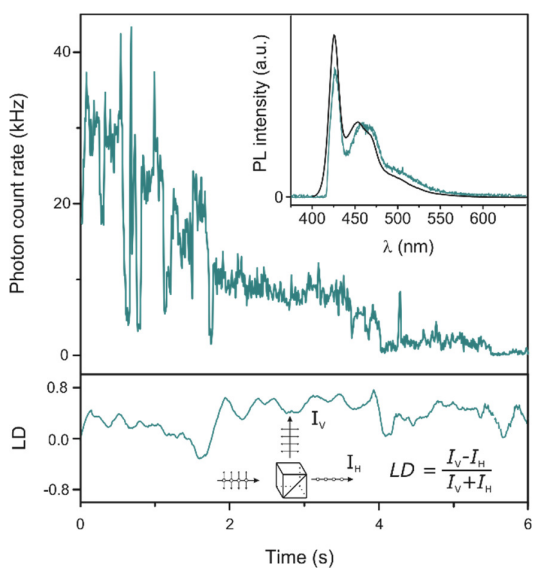

*Figure 2 Representative photoluminescence transient of an individual MSW, showing multistep bleaching, and the corresponding linear dichroism in emission (LD) trace. For the emission LD, the fluorescence light is split by a polarising beam splitter and the LD is calculated as LD = ($I_V$-$I_H$)/($I_V$+$I_H$). The LD trace depicts the temporal evolution of the emission polarisation, changes in the LD correlate with bleaching events in the photoluminescence transient. Inset: Ensemble emission spectrum in solution (toluene, excitation at 400 nm, black line) and two representative single-molecule spectra (in PMMA matrix, thin blue lines). Single molecule spectra were excited at 405 nm and the fluorescence passed through a 409 nm long-pass filter, slightly cutting off the left onset of the spectra.*

We identified the *LD* values for 1683 single MSW molecules, and the results are summarised in figure 3. The *LD* = 0 interval for excitation of the MSWs (defined as the *LD* range from -0.1 to +0.1) shows a distinct population, as one would expect from a molecule containing 12 chromophores arranged in different orientations within a plane. The emission *LD* values also peak in the *LD* = 0 interval, but the distribution is somewhat broader. Broadening of the histogram in emission with respect to excitation can result from stochastic photoinduced exciton localisation[23,33] or bleaching of individual chromophores. In this case, not all chromophores contribute equally to light emission, resulting in a preferential polarisation of emission.

To probe the question whether only one or multiple chromophores radiate at a given instant, we applied time-correlated single-photon counting (TCSPC) in combination with a Hanbury Brown-Twiss setup to the MSW. The sample was excited with a pulsed laser and the fluorescence was passed through a 50:50 beam splitter onto two photodiodes. The photon arrival times on both photodiodes were recorded and the two detection channels were cross correlated to provide the photon coincidence probability as a function of delay time $\Delta\tau$ between the two detectors. Figure 4 shows the $\Delta\tau$ histogram for 281 single molecules, where only the photons from the first intensity level in the PL transient were taken into account, i.e. when the molecule is brightest. The normalised coincidence counts are given for the respective delay time $\Delta\tau$ in intervals with ±0.05 µs width and in 0.1 µs steps (arising from the 10 MHz repetition rate of the pulsed laser). The histogram can be described in terms of the ratio of coincidence photon counts (at $\Delta\tau$ = 0 µs) to "lateral" photon counts at $\Delta\tau \neq 0$ µs, $N_c/N_l$ = 0.27. This pronounced dip at the $\Delta\tau$ = 0 µs interval proves distinct PA in the emission of the multichromophoric MSW. The expected $N_c/N_l$ value for any signal-to-background ratio in the measurement can be calculated as a function of the number of independent emitters.[34]

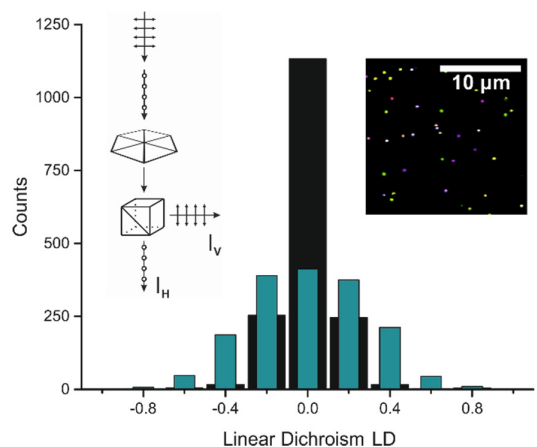

*Figure 3 Histogram of the linear dichroism (LD) distribution for 1683 MSWs for excitation (black bars) and emission (blue bars). Note that the histogram binning is identical for both distributions, but the bars are narrowed for clarity. Left inset: schematic representation of the LD measurement; right inset: fluorescence microscope image of a confocal scan of 20 x 20 µm² (molecules depicted in false colours according to their polarisation in excitation and emission, see text).*

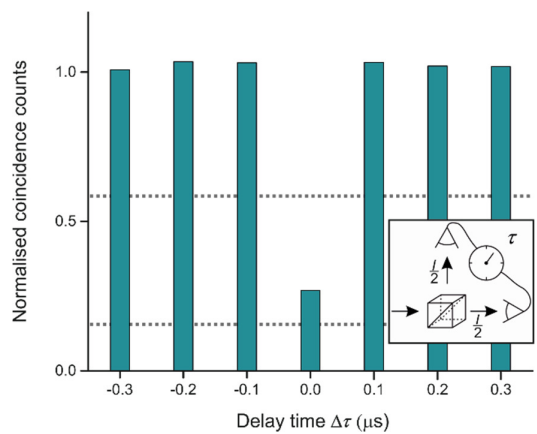

*Figure 4 Histogram of photon coincidences on the two detectors at either side of a beam splitter in the fluorescence path of the microscope as obtained by time correlated single-photon counting of 281 molecules. The dip at the $\Delta\tau$ = 0.0 µs interval, which drops to 27 % of the lateral average, arises due to photon antibunching in the fluorescence of the single MSWs. The dotted lines mark the expected values for one (lower) and two (upper) independent fluorophores. Note that the histogram binning is 0.05 ms but the bars are narrowed for clarity.*

For the present signal-to-background ratio of 23:1, the expected $N_c/N_l$-value for one independent emitter is $N_c/N_l = 0.15$, and $N_c/N_l = 0.58$ for two emitters, as indicated by dashed horizontal lines in figure 4. The measured coincidence rate therefore closely matches the value expected for one single emitter: the MSW with its 12 chromophores behaves as though it were one single chromophore.

PA can only arise in such a multichromophoric system if efficient energy transfer occurs between all chromophores, resulting in singlet-singlet annihilation (SSA),[35,36] the mutual annihilation of two or more excited states. Clearly, however, if emission really did only originate from one single chromophore within the MSW, the *LD* histogram in emission (figure 3, blue bars) would have to show signatures of single-dipole emission, *i.e.* a broad distribution with peaks at the *LD* = ±1 interval.[37] SSA implies that even the most remote chromophores in the MSW must interact with each other. High exciton mobility within the whole MSW is a mandatory prerequisite to make SSA happen before fluorescence from more than one excited chromophore can occur. The existence of PA therefore not only makes dipole-dipole interactions between the chromophores visible but furthermore shows that energy transfer must occur between all 12 chromophores in the molecule. Finally, the fact that the *LD* values in emission tend to zero (figure 3) implies that more than one chromophore within the MSW template must contribute to luminescence. Light emission must therefore arise due to rapid non-deterministic switching between different individual chromophores within the template during the acquisition time of the *LD* value. In other words, the MSW undergoes the process of spontaneous symmetry breaking every time it is photoexcited: excitation and emission occur independently in each excitation cycle, and the emission polarisation becomes scrambled.

In conclusion, we have demonstrated PA in a giant fluorescent π-conjugated MSW of 12 nm diameter. To the best of our knowledge, this structure contains the largest lateral distribution of chromophores for which PA has been observed to date in a covalently bound, structurally defined compound. Effectively, chromophores can couple to each other over distances of almost ten nanometres, *i.e.* excitation energy is effectively transported over at least this length scale. We speculate that in even larger highly ordered templated structures, which are yet to be synthesised, diffusion lengths could be much greater.

Collaborative funding by the Volkswagen Foundation is greatly appreciated. The authors acknowledge support from the European Research Council through the Starting Grant MolMesON (305020). Financial support by the Fonds der Chemischen Industrie and the DFG (SFB 813) is gratefully acknowledged.